# Practical scalability assesment for parallel scientific numerical applications

Natalie Perlin[1,2,3], Joel P. Zysman[2], Ben P. Kirtman[1,2]

[1] - Rosenstiel School of Marine and Atmospheric Sciences, University of Miami, Miami, FL, 33149
[2] - Center of Computational Science, University of Miami, Coral Gables, FL, 33146
[3] - nperlin@rsmas.miami.edu

*Abstract*—The concept of scalability analysis of numerical parallel applications has been revisited, with the specific goals defined for the performance estimation of research applications. A series of Community Climate Model System (CCSM) numerical simulations were used to test the several MPI implementations, determine optimal use of the system resources, and their scalability. The scaling capacity and model throughput performance metrics for N cores showed a log-linear behavior approximated by a power fit in the form of $C(N)=b \cdot N^a$, where a and b are two empirical constants. Different metrics yielded identical power coefficients (a), but different dimensionality coefficients (b). This model was consistent except for the large numbers of N. The power fit approach appears to be very useful for scalability estimates, especially when no serial testing is possible. Scalability analysis of additional scientific application has been conducted in the similar way to validate the robustness of the power fit approach.

*Keywords*— scalability analysis; performance analysis; parallel computer applications; scientific numerical modeling

## I. Introduction

The multi-processor and multi-core era in computer technology and advancements in interconnection networks brought new challenges for software development, opening the possibility of using multiple hardware resources by the same application in parallel. Software performance is no longer dependent only on computer power and CPU speed, but rather on combined efficiency of the given system from both hardware and software perspectives. What became crucial for the performance was effective parallelization of the numerical code, resource management and workload distribution between the processors, interprocessor communications, data access and availability for the processing units, and similar procedures involved in the workflow.  This qualitative progress in computing made many earlier out-of-reach numerical scientific problems doable and enabled further scientific advancement and exploration.  New opportunities that emerged for businesses, the banking industry, medical services, brought new radical networking and teleconferencing solutions, stimulated technological and software advancements in the area of scientific computing as well.

As a consequence of the new commercial opportunities, software development rapidly shifted towards multi-processor applications, multiple-access databases, and advanced communication strategies between the resources. Performance evaluation of such parallel systems became a challenging task, and new ways to quantify and justify the use of additional resources were explored. New terms such as "performance metrics", "scalability", "efficiency", "speedup", scale-up" emerged, followed by numerous studies to find more rigorous and wider definitions to determine methods and tools for their quantitative evaluations [1]-[12].  Out of all new terms, the term *scalability* is likely the most controversial (see, for example, [5], [13]-[16]; and others), which may likely be attributed to different performance goals relevant for the parallel application systems in various spheres. As an example, the business and networking system may need to ensure system stability with the increasing load; commercial application has to assess performance gain a to justify additional hardware resources; or scientific numerical application resource usage needs to be estimated prior to launching a series of experiments. As a wider general definition, *scalability* is the ability of the system or network to efficiently integrate more resources to accommodate greater workload, increase the output or productivity of the system, or enhance its reliability, in proportion to the added resources.

Our current study aims to define specific objectives of performance evaluation in the area of parallel scientific computing, in attempt to determine the optimal use of the available resources. The existing evaluation methods and scalability models are reviewed, and a new empirical model of scalability assessment is suggested to help meeting these objectives.

## II. Key terminology in Performance Evaluation

### A. Performance Metrics

A parallel system involves multi-level parallelism, i.e., a combination of multi-core or multi-processor computer architecture, parallel software algorithms implementations, and interprocessor communication tools such as threading and message-passing. A variety of metrics have been developed to assist in evaluating the performance of the system, and the following will be used in the present study: *speedup, model throughput, scaleup*, and *efficiency*.

- Speedup $S(N)$ is a measure of time reduction, i.e., a ratio of sequential execution time $T(1)$ to that using parallel system of the fixed-size workload (fixed-size speedup), $T(N)$ for N processors as follows [2]:

$$S(N) = \frac{T(1)}{T(N)} \quad (1)$$

- System throughput $Y(N)$ is a system performance metric, reflecting the amount of relevant work done by the

system of *N* processors per unit time., e.g., transactions per second, or forecasted days per second, or simulated time per second of system computation.

- Scaleup C(N), or scaling capacity, is a ratio of model throughput for a multi-processor system to that for the sequential case:

$$C(N) = \frac{Y(1)}{Y(N)} \quad (2)$$

- Scaleup is closely related to the speedup, except it measures the ability of a greater number of processing nodes to accommodate a greater workload in a fixed amount of time (fixed-time scaleup).

- Efficiency E(N), defined as speedup normalized by a number of processors:

$$E(N) = \frac{S(N)}{N} \quad (3)$$

In the ideal case, the efficiency would be equal to one.

### B. Scalability estimates of the parallel system

Scalability refers to the ability of the parallel system to improve the performance as hardware resources are added. In contrast to the terms in the previous Section, for which a quantitative definition is straightforward to establish, the term "scalability" is more controversial, with numerous efforts in search of a rigorous definition [3], [5], [8], [13], [16]-[18]. One of the definitions that captures the essence of the term is given by [6]: "*The scalability of a parallel algorithm on a parallel architecture is a measure of its capacity to effectively utilize an increasing number of processors*". The lack of scientific precision of this definition is due to a variety of factors within a given parallel system affecting the performance that need to be considered.

Reference [8] offered an analytical definition of scalability, as well as few more scalability metrics. According to their definition, scalability is the ratio between the two efficiency estimates, such that

$$SC = \frac{E(N_2)}{E(N_1)}, \text{ for } N_2 > N_1. \quad (4)$$

Ideal parallel system scalability will be equal to one. This could be another performance metric for numerical analysts to consider. In most of the literature, however, scalability is evaluated in a wider sense. Scalability is generally considered to be a system property [17], in which a number of independent factors (i.e., scaling dimensions) could be manipulated to some extent. The key factor to achieve scalability for a particular application could be a tradeoff between the speedup and efficiency.

Studies of parallel computer systems performance range from discussions of possible scalability models and metrics [2], [4], [6], [19] to methods of scalability analysis for a system of interest, to methods of scalability prediction [7], [10]-[12], [14], [15], [20]. Scalability models are developed to predict system performance as hardware resources (processor cores, or cpu-s) are added to the system. Scaling capacity is often used as a performance metric. Table 1 shows most common scalability models, which are subsequently illustrated in Fig. 1.

TABLE I. COMMON SCALABILITY MODELS.

| Model | Scaling capacity estimate | New parameters | Refs. |
|---|---|---|---|
| Linear | $C_L(N) = N$ | $N$- processor count | ideal case |
| Amdahl's Law | $C_A = \frac{N}{1 + \sigma(N-1)}$ | $\sigma$ - a fraction of work that is sequential (0-1), or seriality parameter | [21] |
| Gustafson's Law | $C_G = \frac{N^2}{N + \sigma'(1)(N-1)}$ | $\sigma'(1)$ – fraction of serial work with one processor | [19], [22] |
| Superserial Model | $C_S(N) = \frac{N}{1 + \sigma[(N-1) + \gamma N(N-1)]}$ | $\gamma$ - fraction of serial work for interprocessor communications (0-1), or superseriality parameter | [4] |

The Linear model represents an ideal case, in which the scaling capacity equals to or is directly proportional to the number of processors (hardware resources) used. With the few exceptions, realistic applications would not achieve linear behavior, and will have losses due to code seriality, inter- and intra-processor communications, network speed or other resource or data exchange limitations, which result in sub-linear system performance.

The Amdahl's Law [21] model assumes the sequential fraction of the code, $\sigma$, is fundamental limitation of the speedup of the system, and thus the scaling capacity would asymptotically approach $(1/\sigma)$ value. However, in case of a faster processor being used, both serial and parallel parts of the code would experience speedup to some degree. Amdahl's Law also assumes that the parallel fraction of the application, $(\pi = 1 - \sigma)$, remains invariant of the number of processors. In some applications, however, the amount of parallel work increases with the number of processors, while the amount of a

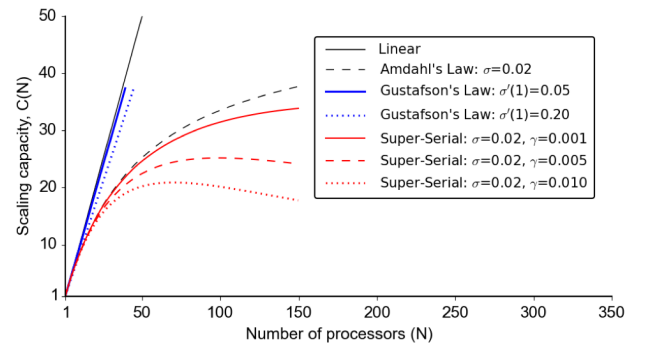

Fig. 1. Scalability model examples: Linear, Amdahl's Law, Gustafson's Law Super-Serial Model (See Table 1 and Section II.B).

11/ 2016

serial work, $\sigma'(1)$, remains constant. This case is described by Gustafson's Law, also called scaled speedup (see Table 1; [19], [21]). In other examples, applications with very large spatial domain require domain decomposition and may not run on a serial processor due to memory constraints, and thus scaling capacity (as in Section II.A) is not technically possible to estimate. Super-Serial Model ([4], [23], [24]) takes into the account the fraction of serial work used for interprocessor communication ($\gamma$). This additional factor has an important effect on the scaling capacity profile, resulting in a performance roll-off past a scaling capacity maximum at $N_C$ number of processors, $N_C = \sqrt{(1-\lambda)/\lambda\sigma}$, beyond which adding more processors becomes counter-productive.

### C. Scalability of Parallel Scientific Applications

Many of the modern scientific applications for parallel systems are complex software packages or existing code models, which are applied for certain numerical experiments, and often run on a shared resources at larger supercomputing centers. With limited or no control over the software algorithms, code parallelism, or hardware resources limitations, the major goal for the user of such a parallel system would be to maximize the cost-effective use of computational resources for a given application. Scaling dimensions that influence performance and scalability of a given parallel system are associated with the following:

- Message-passing interface (MPI) implementation;
- Particular application options affecting the choice of algorithms;
- Compilers available and optimizations utilized by the software application;
- Methods of spatial grid decomposition of a modeling domain for multi-processing, i.e., domain topology (maximum or minimum grid tiles, pre-determined spatial grid decompositions, etc.);
- Additional layers of abstraction or third-party packages for data exchange between the model components in multi-model coupled systems (Model Coupling Toolkit/ MCT; Earth System Modeling Framework/ ESMF; etc);
- Number of CPUs available for a particular user (in a shared-resources machines) and other imposed CPU-time limitations.
- Queue and job scheduling parameters, computer system load.

This is far from a complete list of possible independent scaling dimensions, but it indicates that the problem needs to be approached from a practical side, rather than theoretical or analytical. From a single user's perspective, the scaling analysis of one's parallel scientific application is needed to determine *the most efficient balance* between CPU-time required, number of processors or cores used in a given platform, wall-clock time to perform calculations, memory demands, possible computer system workload and queue waiting time. Inefficiency could be costly not only to the users of such applications, but to the users and owners of the shared computer system or computer center.

Excessive and unnecessary power consumption when more processors than efficiency would require are requested, reduced system availability to the other users, and creation of long-waiting queues are only few of the issues. A simple and quick assessment method of scalability of a given parallel application is therefore needed to determine the optimal processor/core count, particular queue choice, etc. Section 3 presents some case studies and their performance analysis, and suggests such an assessment method based on several short pre-production test runs.

### III. PARALLEL SCIENTIFIC APPLICATION STUDIES

#### A. Climate Model Case Studies

We performed several simulations of the Community Climate Model System (CCSM) using the supercomputer resources at the University of Miami's Center of Computational Science (CCS), and conducted a performance analysis. This assessment included different MPI implementations, and evaluation of the scalability of the model applications with increasing computational resources. The CCS supercomputer is based on IBM iDataPlex dx360 M4 systems (totaling 5500+ computational cores) with Mellanox FDR InfiniBand interconnect. Each computational node has 16 cores (two 8-core SandyBridge Intel 2.6 GHz processors) and 32GiB of memory (2 GiB per core).

CCSM is a coupled general circulation model covering the entire globe, and is aimed to simulate and predict climate [25], [26]. This modeling system combines several model components with global coverage: atmospheric model (extending vertically from the surface to the top of the atmosphere), ocean model (extending down from the ocean surface to the bottom), land model, and ice model. The information exchange between the models at regular intervals allows one model's output to be used for other models' calculations and boundary conditions. The standard code of one of the most recent CCSM version, CCSM4_0_a02, and two types of standard model grids were used in the presented tests. The low resolution grid 0.9x1.25_gx1v6 with dimensions of 288 x 192 points in longitudinal and latitudinal directions of the atmospheric domain, respectively, and the global ocean model grid sizes 320 x 384 points, will be referred to as "LowRes" test. High resolution grid 0.47x0.63_tx0.1v2 with dimensions of the atmospheric grid 576 x 384 points, and that of the ocean grid 3600 x 2400 points, will be referred to as "HighRes" test. The land model follows the atmospheric model resolution, and the ice model follows the ocean model resolution.

All the tests included 5-day forward model integration, from the cold start. The usual length of the production forecasts are months or years, so that 5-day forecasts are considered short pre-production test runs.

One computational node of the cluster includes 16 cores, and LowRes simulations always used all cores of the computational node, when number of processors (cpu-s) was 16 or greater. LowRes tests were performed using a number of processors from 1 to 1024, doubling the number of processors in each successive series of tests (i.e., 1, 2, 4, 8, 16, ..., 1024). For the tests using the openmpi implementation, the highest number of cpu-s to successfully complete the test was 512. (Note that the terms "cpu" and "core" are often used interchengeably further in text.)

Due to high memory demand and the large size of the domain in the HighRes simulations, the lowest number of cores to successfully complete the test was 32; number of cpu-s doubled in each successive test, up to 2048. The tests involving smaller number of processors have been done with the following computational resources reservations: 64-cpu runs reserved 4 cpu-s per node, and 32-cpu runs reserved 2 cpu-s per node; in all cases 28GB of memory per node was reserved for the tests.

### B. OpenMPI and MPI Implementations Analysis

Intel Composer_xe_2013.2.146 compiler suite with *ifort* and *icc* compilers was used to configure several implementations of Message-Passing Interface (MPI), in particular, OpenMPI (lower-case *openmpi* is used to reflect a specific module name for the versions *1.6.2*, *1.6.4*, *1.6.4-mxm*, *1.7.5*) and *impi* version *4.1.1.036*, to further be used to compile the CCSM4_0_a02 code. Hyperthreading (HT) was disabled on the processors. Series of six (6) identically set LowRes CCSM tests were run on the same set of 4 computational nodes (64 cores total) to additionally evaluate the extent of variability in the application execution time. Fig. 2 shows the total time used in each test, as reported by the LSF job scheduler at the end of each test. There appears to be some small variability between the tests within each MPI implementatin. Most differences were found between the MPI versions, such that *openmpi/1.7.5* and *impi* consistently yielded about 30% faster model run time than the earlier OpenMPI versions. Based on the results of these tests demonstrating superior performance of *openmpi/1.7.5* and *impi*, only these MPI implementations were be used in subsequents tests.

### C. Climate Model Timing Diagnostics

Some useful timing metrics are a part of the CCSM diagnostic suite reported at the end of each simulation. Total time spent to complete the numerical tests is partitioned into three main phases: the first phase is to initialize the model, the second phase is to integrate model forward in time (main forecast), and the final model cleanup is the termination phase. The final phase took negligible time. Model initialization phase, however, could be comparable to the time spent for the daily forecast in our short-duration tests. The model integration time as a major phase of the numerical test and varies depending on the forecast length. The compute rate, equivalents to the wall-clock time in seconds needed to complete 1 day of forecast, so the resulting units are *s/day*. The more processors are used, the less time it takes to complete the forecast of a given time; it could also be used to compute the speedup using Eq. (1).

Two diagnostic timing variables, model initialization time and compute rate, averaged from the series of LowRes CCSM tests performed using *openmpi/1.7.5* and *impi* implementations, are shown in Fig. 3. Compute rate for the LowRes tests indicated steady speedup with the increased number of processors, from over 1500-1600 s/day on a single processor, down to about 22 s/day (*openmpi/1.7.5* using 256 cpu-s) and 15 s/day (impi using 512 cpu-s). Further increase in computational resources did not produce any speedup, and sometimes performance retrograded. The initialization time (Fig.3, right y-axis) initially decreases from over 110s for 1-cpu tests to lowest values of 18s and 14s for the tests using 64 cpu-s with *openmpi/1.5.7* and *impi* implementations, respectively. Initialization time increased in all the test involving greater than 64 processors. It also worth noticing that for tests involving 256 cpu-s or greater, the initialization time exceeds the model compute rate. For the accurate performance analysis it is therefore necessary to separate the initialization time from the main forecast execution time in computed performance metrics.

### D. System Usage Breakdown Analysis

The increased number of processors used in simulations speeds up the computations, but also increases the load of message-passing data exchange between the processors. Fig. 4 demonstrates the breakdown analysis (also called "bottleneck" analysis) of total time (100%) spent for running the parallel application into contributions from the major tasks. These three major tasks that could be performance-limiting are the following: time spent for computations *per se* using CPU resources (CPU), time spent for message-passing calls (MPI), and time spent for input/output (I/O). This analysis was made available by using Allinea Tools© PerformanceReport software; the tests involved *openmpi/1.7.5* and *impi*, and used from 1 to 64 cpu-s (a limitation imposed by Allinea software license). In all the tests, actual application operations and computations took most of the time, so that the tests were "CPU-bound". Time spent for MPI calls increases rapidly with the number of processors, reaching about 37% of total application run time in 64-cpu tests. Input and output operations consumed 0.1%-1% of the total time. Use of message-passing (MPI) enables speed up of the computational task by spreading the calculations into a greater number of processors, but MPI communications present an overhead expense. Exact correspondence between CPU – MPI – I/O strictly depends on a particular application and its setup on a given compute system. For the LowRes CCSM tests shown in Fig.4, time spent for MPI calls increases rapidly with the cpu count. If time for MPI calls exceeds the time spend for application operations, it is said to be an "MPI-bound". The performance of an application further decreases with the increased number of processors due to the excessive overhead of MPI communications.

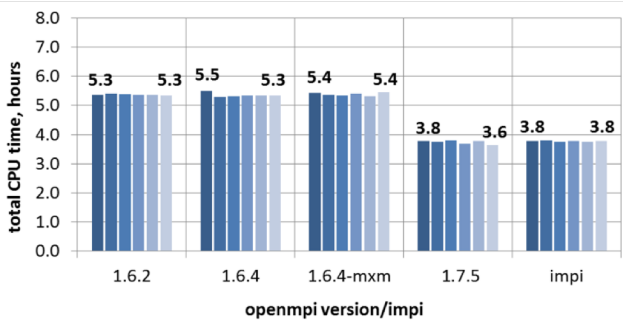

Fig. 2. CPU-time reported for the CCSM simulations, compiled with the Intel Fortran compiler (ifort) but employing different MPI implementations. All the tests were performed using 64 cores involving 4 nodes (16-core nodes), and series of 6 tests were conducted for the each MPI version. See text in Section 3 for more details on the CCSM run settings. Numbers above first and the last column in each group of tests are hours of total cpu-time, given for illustration purposes.

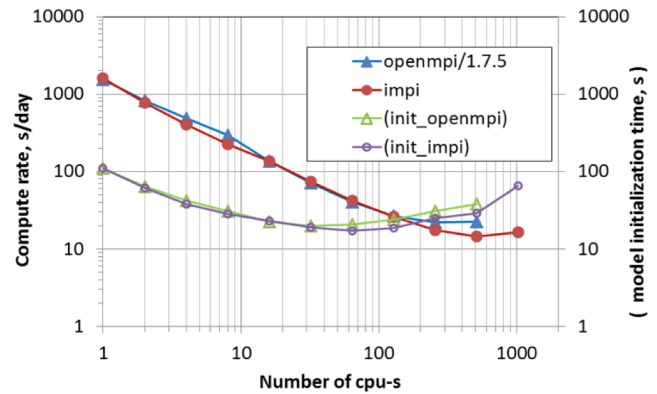

Fig. 3. CCSM diagnostics of compute rate (left y-axis, blue and red lines corresponding legend entries with no parentheses) and initialization time (right y-axis, green and purple lines, legend entries in parentheses), for the two MPI implementations, *openmpi/1.7.5* and *impi*. Note the logarithmic scale on all axes. All values except for 1024 cpu-s, are averages for the series of 6 identically set up tests. No confidence intervals or standard deviations were shown due to a minimal variability.

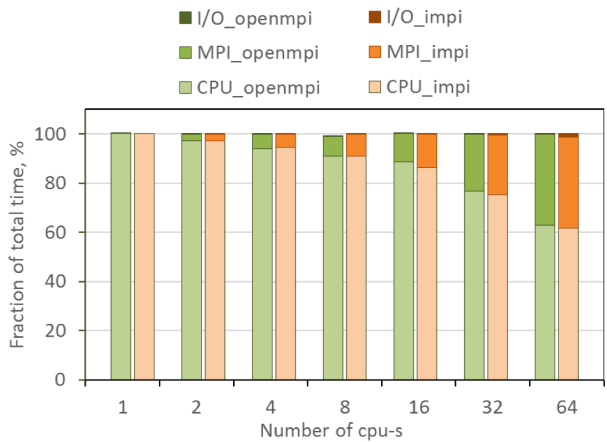

Fig. 4. Breakdown of the total time of a simulation (100%) into the percentage of time spend for numerical calculations (CPU, lightest shading), message-passing communication (MPI, medium shading), and input/output data reading/writing (I/O, darkest shading), based on the Allinea Software © Performance Reports. Tests were conducted for *openmpi/1.7.5* (left columns) and *impi* (right columns), and using 1 to 64 processors. CPU-Results reported are averages of the series of 6 tests per each combination.

### E. Numerical Application Scalability Analysis

Evaluating a scalability of our parallel application on a multi-processor parallel system started from estimating the scaling capacity metrics $C(N)$ (Eq. 2) from the ratio of model throughputs $Y(N)$ and $Y(1)$, resulted from the simulations using $N$ processors and one processor, respectively. CCSM model throughput is estimated in units of simulated years per day of computation time. The model throughput grows with the number of computing resources, and it is used to estimate the scaling capacity (scaleup, as in Eq.2). It is closely related to the model compute rate (shown in Fig.3), except the compute rate is used to estimate the speedup (Eq.1). The scaling capacity for LowRes simulations is shown in Fig. 5. Because of the number of cpu-s ranging over three orders of magnitude (from 1 to 1024), it is convenient to plot the results on the logarithmic scale. Note that when both axes implement logarithmic scale, the scaling capacity shows as nearly a linear behavior up to the upper range of $N$, after which the performance declines. This linear behavior naturally fits a power law function for $C(N)$:

$$C(N) = b \cdot N^a, \quad (5)$$

where $b$ and $a$ are empirical constants. While $b$ is a dimensionality coefficient, $a$ is a linearity coefficient. The closer $a$ and $b$ are to 1, the better the fit is to linear, and will be sub-linear for $a<1$. Such straight line fitting for LowRes tests (with *impi*) for the range of 1–512 cpu-s was calculated using the least squares method, in particular, using Microsoft Excel *LINEST* function. The results showed $b=1.28$ and $a=0.77$, as indicated in Fig.5 legend as well.

Alternatively, we could consider an application when it is not possible to conduct a serial experiment, such is the case with CCSM HighRes test. No throughput data is available for for $N=1$, and thus scaling capacity $C(N)$ could not be estimated using Eq.(2). Fig. 6 shows model throughput $Y(N)$ from the LowRes and HighRes simulations (non-normalized, on the left y-axis), as well as the model cost estimated as processor-hours per year of forecast (solid lines, right y-axis). Because of the higher resolution and larger physical model domain, HighRes tests have notably lower throughput and higher model cost. Similarly to the scaling capacity curve in Fig.5, the logarithmic scale demonstrates that the throughput $Y(N)$ has a linear profile except for the highest $N$, i.e. $Y(N)$ can be approximated with the similar power fit:

$$Y(N) = b \cdot N^a \quad (6)$$

The coefficient $b$ no longer is expected to be close to 1, as it is a dimensionality coefficient for $Y(N)$ as opposed to dimensionless $C(N)$. The power coefficient $a$, however, is still a strong indicator of linearity. The calculations of the power fit are done in a similar way using Microsoft Excel. Power coefficient $a$=0.77 for $Y(N)$ in the LowRes tests is equivalent to the corresponding coefficient for $C(N)$ to the sixth decimal digit, which indicates that the power fit is a good alternative to evaluating the corresponding scalability. Power coefficient is higher for HighRes tests, $a$=0.91, indicating that the particular application scales better with a greater number of processors. Because the computer system is identical in the LowRes and HighRes simulations, only the differences in the application configuration determine its scalability on a given system.

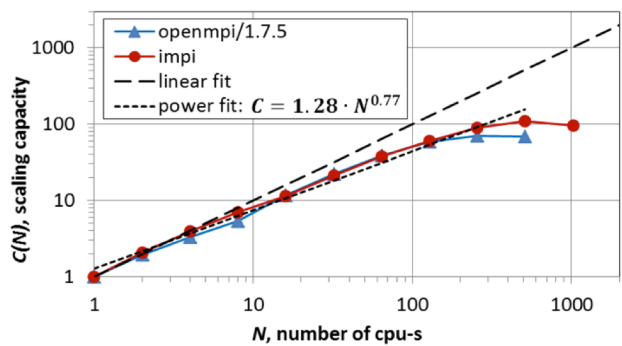

Fig. 5. Scaling capacity estimated as normalized model throughput, from Eq.(2), for CCSM LowRes tests with two MPI implementations. Note the logarithmic scale on both axes. Long-dash line marks linear scaling, and short-dash marks a power-fit line as indicated in the legend, as a function of number of processors *N*. In all the tests except for *N*=1024 cpu-s, throughput was averaged for 6 identically set simulations.

Differences between MPI implementations become apparent in LowRes tests performance for cores of 256 and higher (Fig. 6), when simulations with *impi* produced higher model througput than those using *openmpi/1.7.5*; LowRes test using 1024 cores and *openmpi/1.7.5* was not able to complete (due to suspected I/O issues). It also worth noting that model costs grows moderately slow with increasing number of processors at the lower cpu count, but a greater increase rate occurs at higher cpu counts, i.e, for $N$ =512,1024 in LowRes tests, and $N$ =2048 in HighRes tests. This is consistent with the lower throughput produced in corresponding simulations.

In summary, the proposed power law fitting method of model performance metrics from shorter pre-production tests allows quick practical assessment of the application scaling in a given parallel system. The cpu-count at which the metrics start to deviate from a linear fit on a logarithmic scale would give an indication of the performance saturation, beyond which scaling is not efficient.

*F. Alternative Scaling Estimates*

We considered alternative methods to estimate the scalability of the CCSM simulations as well. It has been shown earlier that a Super-Serial Model expects the system performance to peak at a certain critical cpu-number, declining with furter increase in cpu (Table I) similar to the behavior found in CCSM LowRes tests. We used a least squares method to find a fit of the scaling capacity from the CCSM tests with impi to a custom function described by the Super Serial Model. Python package *LMFIT* was used for that purpose, to determine the coefficients $\sigma$ and $\gamma$, which were searched in the range between $10^{-5}$ to 0.5; the fitting curve to the $C(N)$ data is presented in Fig. 7. The fitting coefficients were $\sigma = 6.85 \cdot 10^{-3} \pm 5.68 \cdot 10^{-4}$, and $\gamma = 3.13 \cdot 10^{-4} \pm 1.44 \cdot 10^{-4}$; critical processor numer $N_C$ was 683 cpu-s. Note that the SuperSerial model fit to the original data smoothed the performance peak observed at 512 cpu-s. Attempts to synthesize additional data by linearly interpolating the $C(N)$ to the entire range of integer cpu-numbers from 1 to 1024 (for 100 to 1024 cpu-s, adding data to the upper range only) did improve the confidence intervals of the coefficients, $\sigma = 6.77 \cdot 10^{-3} \pm 3.86 \cdot 10^{-5}$ ($6.75 \cdot 10^{-3} \pm 1.58 \cdot 10^{-4}$); $\gamma = 3.23 \cdot 10^{-4} \pm 1.00 \cdot 10^{-4}$ ($3.25 \cdot 10^{-4} \pm 4.27 \cdot 10^{-5}$); which negligibly affected the fitting curve shape, and the peak at 512 cpu-s was still missed by the functions. Critical $N_C$ was equal to 676, and 675 cpu-s, respectively. Two addiational model fits are not shown in Fig. 7 due to minimal differences with the original fitting solution. It appear that finding a fit for a customized model could be more unwieldy than a straight line fit for the more standard power law function.

An alternative definition of scalability $SC$ from Eq.(4) was also considered, for the same group of tests (Fig.7). The efficiencies $E(N)$ were computed as in Eq.(3), except scaling capacity $C(N)$ was used in place of speedup $S(N)$. There is an definite plunge in $E(N)$ and decline in $SC$ parameters that could be used as an indication of the poor performance, and assist in choosing the optimal cpu load. However, this method requires the scaling capacity estimates for the serial similation, which may not be available (as in CCSM HiRes cases).

IV. MOLECULAR DYNAMICS APPLICATION TESTS AND ANALYSIS

We tested an application from a different field of science and conducted similar performance analysis to test robustness of the scalability estimate method using the power law fit proposed in the previous Section. Desmond is a software package developed at D.E. Shaw Research (*DESRES*) to perform high-speed molecular dynamics simulations of biological systems on platforms with a large number of processors [27]-[29]. *Desmond* is integrated with *Maestro* package developed by *Schrödinger, Inc.*, which is the molecular modeling environment, for setting up simulations and graphical viewing. The application version of *Desmond2014.2/maestro* was used in the present tests. This software has its own version of MPI communications to enable multi-processing.

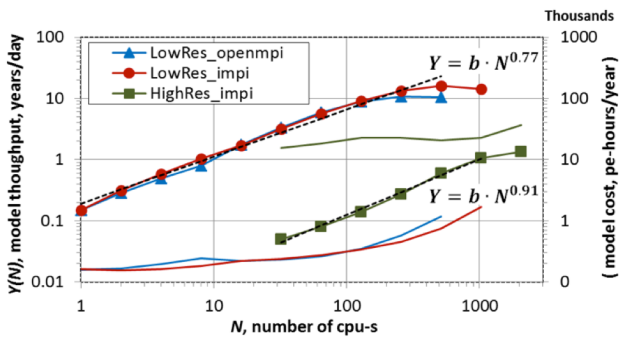

Fig.6. Model throughput (left y-axis, lines with symbols as appear in the legend) and model cost (right y-axis, solid lines using the corresponding colors as in the legend) for the CCSM LowRes tests using openmpi/1.7.5 and impi, and HighRes tests using impi. Note the logarithmic scale on all axes. Dashed lines are power fit for the model throughput *Y(N)* from LowRes and HighRes tests with impi implementations, following Eq.(6). Coefficient *b* equals 0.19 and 1.9*10[-3] for LowRes and HighRes fitted lines, respectively, and power coefficient *a* equals 0.77 and 0.91 for the corresponding lines.

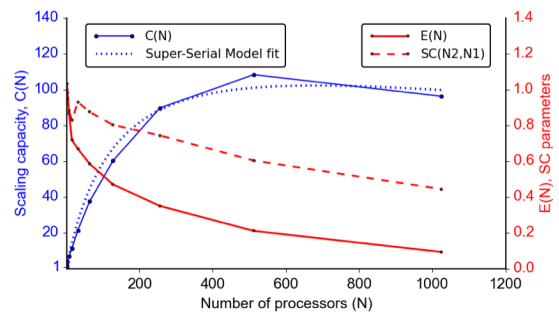

Fig. 7. Additional scalability analysis of the CCSM tests with *impi*. (left y-axis): Scaling capacity *C(N)* and corresponding least-squares fit of the Super-Serial Model. (right y-axis): Efficiency *E(N)* estimated from Eq.(3), and scalability *SC* estimated from Eq.(4), for the same set of tests. See Section III.F for numerical details on the fitting parameters.

The test pre-production simulations to determine the optimal number of cpu-s were conducted for the following configurations: 1, 2, 8, 16, 32, 48, 96, 144, 192, and 256 cores. The total length of molecular dynamics simulation was 1.2 ns. Each simulation contains eight (8) stages of data processing or computations, out of which only stage 8 is computation-bound. Log files contained timing for each stage, as well as the software-estimated model throughput at each time step during stage 8; we used python script to process the log files and derive mean values of the throughput. Note that all of the computation stages shown experience speedup with the growing number of cpu-s, but it is the computationally-bound stage 8 that would be most important for the performance assessment. In this case, model throughput $Y(N)$ has units of ns/day, i.e., ns of simulation per day of computation. Microsoft Office Excel was further used to compute power law coefficient for the model throughput following the Eq.(6) in a similar way as described in the previous Section.

Fig. 8(b) summarizes the findings of the performance and scaling analysis for $Y(N)$, where power coefficient was found to be *a*=0.56, indicating not so optimal scaling of the application overall (ideal case *a*=1). Performance improvement slows down for greater than 48 cpu-s, and stops or declines for greater than 96 cpu-s (when the model throughput curve crosses the power fit line in Fig.8). Upon this finding we communicated with the software developers, who informed us that the target size of domain decomposition for such a simulation would be 32 cpu-s. Our testing indicated that the parallel application could successfully run on 48-96 processors.

V. SUMMARY AND DISCUSSION

The present paper revisits performance and scalability analysis, and determines specific needs and questions to be answered by such analyses in the area of parallel scientific computing applications. These needs are often related to the optimal use of given hardware resources, CPU-time allocated for the project, and wall-clock time to complete the numerical experiments. We performed scalability analysis for the two parallel scientific applications, conducted on the supercomputer in the Center of Computational Science (CCS) of the University of Miami. The computer system is built on IBM iDataPlex dx360 M4 systems, totaling 5500+ computational cores.

First scientific application used for series of numerical simulations was Community Climate Model System (CCSM), a coupled general circulation model covering the entire globe and consisting of several components: an atmospheric model, an ocean model, a land model, and an ice model. The standard code of CCSM4_0_a02 was used in the presented experiments, along with the two configurations of model grids having different horizontal resolutions that were referred to as the "LowRes" and "HighRes" simulations. All the experiments were short pre-production tests with model integration forward for 5 days.

First, several Message-Passing Interface (MPI) implementations based on Intel *ifort* and *icc* compilers were tested to eliminate MPI-based performance limitations. Performance using MPI implementation with Intel impi version 4.1.1.03 was found to be superior of several others; performance of *openmpi/1.7.5*, was comparable to that with *impi* on a lower number of cores, but declined for number of cores *N*=256 and higher. No LowRes test was possible to complete for *N*=1024 using *openmpi/1.7.5*. All HighRes tests were done using *impi* implementation, up to *N*=2048. The breakdown of resource use, or bottleneck analysis made using *Allinea Tools PerformanceReport* software, indicated that the time spent for MPI calls increased rapidly with the growing number of cores in LowRes application, up to 37% of the total time for *N*=64 (the highest number for which the analysis was made possible). Scaling of this application to a greater number of processors is therefore expected to become MPI-bound due to growing overhead from the MPI communications.

Scaling capacity of the LowRes application was further estimated by a normalized model throughput, and was found to be sub-linear. It was demonstrated that the scaling capacity $C(N)$ for the *N*cpu-s could also be fitted with a line representing a power fit in the form of, $C(N) = b \cdot N^a$, where *a* and *b* are constants; the scaling approaches linear when these coefficients approach 1. The scaling capacity started to deviate from this power fit for the large number of *N*, indicative of addition of hardware resources being counterproductive. Model throuput $Y(N)$ on a logarithmic scale followed the same log-linear behavior as the scalability, with the identical power coefficient *a*, but different dimensionality coefficient *b*. Power coefficient *a*=0.77 resulted for the LowRes simulations,, and *a*=0.91 resulted for the HighRes tests. This power fit approach becomes especially useful for scalability estimate when no test is possible to conduct using a single cpu (*N*=1), such as the case with the HighRes simulations due to a large domain size and memory requirements. Application scalability depends both on system configuration and application configurations. Because of our tests conducted on the system with the same configuration, including using the same type of MPI, greater power coefficient indicated a better scalability of the HighRes simulations. When model throughput started progressive deviate from the power line fit, it marked the limit of efficient usability of hardware resources, beyond which performance decreased or retrograded.

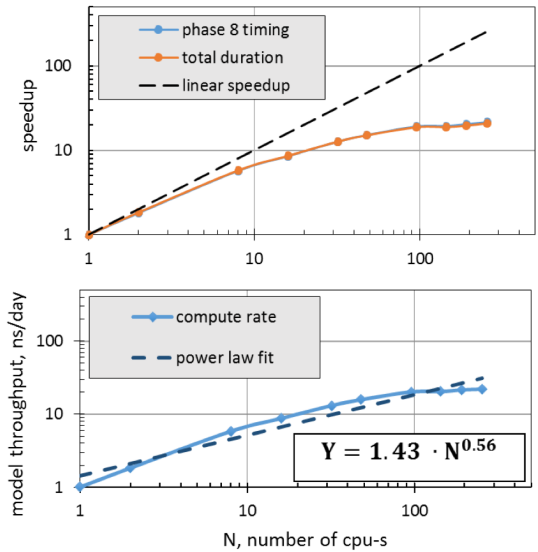

Fig. 8. (top): Average speedup as in Eq.(1) computed for Desmond/maestro simulations for the total simulation duration, and for the computation-bound phase 8 of the experiments. (bottom): Average compute rate reported by the software and the power fit as in Eq.(6).

The robustness of this simple power fit approach of the scalability estimate was further tested on series of additional numerical experiments using an application from a different scientific area of molecular dynamics (*desmond/maestro* application). The power coefficient was found to be *a*=0.56, notably lower than in the previous tests using a climate model. Nonetheless, the optimal number of cpu-s for the given application determined using this method was comparable (and greater than) the number of cpu-s targeted by software developers.

We also considered some other methods for scalability estimate, such as using a Super-Serial model fit, or alternative definition of scalability as the ratio of efficiencies. These methods, however, were either more complicated to find a fitting curve (such as for Super-Serial Model), or requiring prior scalability estimates involving serial test with N=1, which is not always possible to conduct There were some earlier works that suggested a logarithmic approach to performance analysis [15], in particular, to performance prediction. Their technique, however, suggested more analytical way of performance prediction, considering several factors, each with its own log-dependence on the number of processors. This appears to be of a limited use from a point of view of the practical goals defined for our study, and answering the question about the optimal usage of the given computing resources.

As a bottom line, our proposed scaling analysis technique for the parallel scientific application assumes conducting a number of shorter tests, determine most relevant performance metrics (such as model throughput or compute rate), and finding a power fit model. Careful selection of performance metrics, either model-supplied or derived by the post-processing of the output, is the key to the correct evaluation of the longer experiments performance. The power fit approach appears to be robust due to its simplicity (essentially linear fit), applicability to a wide range of system configurations, including those not possible to conduct in a serial mode, and various degrees of sub-linearity of parallel applications.

ACKNOWLEDGMENT

The authors would like to thank the Advanced Computing group from the Center of Computational Science (CCS, University of Miami), in particular, ZongJun Hu, Warner Baringer, and Pedro Davila, for support and guidance on using the computer facilities and solving the issues. We thank Dr. Stephan Schürer and Dr. Dušica Vidović from the CCS/UM for the help with the *Desmond* package. We thank anonymous reviewers for the comments contributed towards improvement of the manuscript.

This research did not receive any specific grant from funding agencies in the public, commercial, or not-for-profit sectors.